# ARGO-SLSA: Software Supply Chain Security in Argo Workflows

Thariq Mohomed
*Department of Computer Science*
*Informatics Institute of Technology*
Colombo, Sri Lanka
seyed.20200758@iit.ac.lk

Indrajith Ekanayake
*Department of Computer Science*
*Informatics Institute of Technology*
Colombo, Sri Lanka
indrajith.e@iit.ac.lk

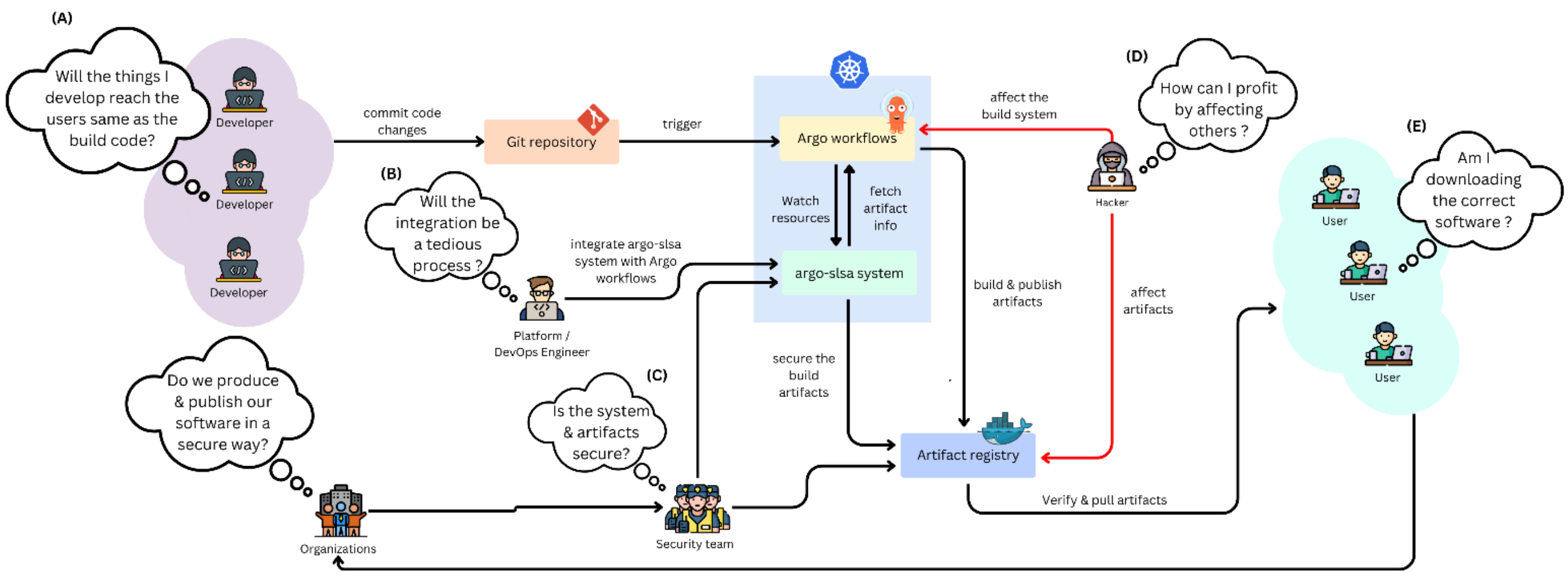

Fig. 1. In this paper, we present ARGO-SLSA, a system that allows organizations to (A) implement software supply chain security best practices (B) in an easily Integrable system onto Argo workflows (C, D) to mitigate external threats to the software products (E) providing the ability to deliver secure & traceable artifacts to the consumers.

***Abstract*—Kubernetes has become the de facto standard when it comes to managing microservices. Automating complex, multi-step workflows is a common requirement in Kubernetes. Argo Workflows is a Kubernetes-native engine for managing these workflows in an automated fashion. These workflows generate artifacts such as executables, logs, container images, and packages. These artifacts require proper governance. Open-Source Security Foundation (OpenSSF), in collaboration with Google, introduced Supply-chain Levels for Software Artifacts (SLSA), a security governance framework that includes detailed technical requirements for producing artifacts. However, Argo Workflows doesn't have any built-in ways to provide the ability to incorporate the SLSA framework. This vacuum creates silos because practitioners need to rely on third-party tools to meet software supply chain security standards. This paper proposes a Kubernetes-native controller written to run in parallel to the existing open-source Argo Workflows to enhance the security of artifacts. Cryptographic signing and provenance attestations for the artifacts produced by the controller, which allows Argo Workflows to comply with SLSA standards. Evaluations were conducted in a real-world, self-hosted environment to demonstrate ARGO-SLSA's ability to elevate artifacts to Level 2 of the SLSA compliance build track. Experimental results indicate that the ARGO-SLSA controller surpasses existing software supply chain security solutions.**

## I. Introduction

The inherent requirement for high maintainability and scalability in software is frequently addressed through the adoption of microservice architecture, which is distinguished by its key features, such as bounded context, flexibility, and modularity [1]. This architectural style typically necessitates the decoupling of services into independent containers, which are then orchestrated via an engine such as Kubernetes, Docker Swarm, and OpenShift. Among them, Kubernetes archived mass adoption due to customizable features, mature security features, flexibility, scalability, and reliability [2]. Complex Kubernetes batch processes such as machine learning (ML) workflows [3], data processing workflows [4], [5], and continuous integration/ continuous delivery (CI/CD) pipelines [6] require workflow automation engines such as Argo Workflows [7], Apache Airflow [8], and Tekton Pipelines [9]. These workflows often generate container-native software artifacts such as container images, Helm charts, open policy agent (OPA) bundles, and software bills of materials (SBOMs) [10].

Argo Workflows is a popular open-source cloud-native workflow engine for orchestrating parallel jobs on Kubernetes [7]. Even though it is a strong tool that can leverage the highly scalable cloud-native model for workflow automation, it doesn't provide a native way to secure the artifacts built through its system. This leaves the responsibility of securing the supply chain in the developers' hands; this can lead to many issues, such as inconsistent security practices and lack of standardization, loss of developer productivity by focusing on security practices, delayed detection can lead to huge damage to both the artifact producers and consumers, failure to meet industry standards can lead to compliance issues, and lack of artifact traceability without proper provenance can reduce trust and accountability [6], [11], [12]. However, modern CI/CD systems like Google Cloud Build [13], GitHub actions [14], and workflow automation tools like Tekton Pipelines [15] all provide ways to achieve software supply chain security through Supply-chain Levels for Software Artifacts (SLSA) framework [16] compliance for all its users without much involved implementations around software supply chain security area [16]. Hence, there's a pressing need to address this gap in Argo Workflows with an efficient, easy-to-integrate, and scalable solution to secure its build artifacts. This paper addresses the following research objectives:

**RO1:** *To secure artifacts without modifying the existing Argo Workflows Operator implementation*

**RO2:** *To mitigate reconciliation conflicts.*

**RO3:** *To provide SLSA compliance to artifacts created with Argo Workflows*

The remainder of this paper is organized as follows. Section II reviews the relevant literature surrounding the problem domain, as well as the frameworks used for the proposed validation model. Sections III and IV detail the methodology and evaluation of the proof-of-concept, respectively. Section V synthesizes the conclusion and explores potential future directions in software supply chain security. Finally, Section VI includes how to obtain the complete source code for the proposed solution.

## II. Related Work

Automated building, testing, and deployment in the cloud through CI/CD have revolutionized software development, and their adoption has accelerated rapidly [17]. While this shift provides a great number of advantages, if any software artifact produced by these systems is not properly secured throughout the supply chain, it will introduce obvious vulnerabilities, exploit software consumers, and impact the software producers' credibility [2]. This is evident in high-profile incidents like SolarWinds [18].

When it comes to the topic of software supply chain security, there's a considerable amount of work that has been done and standardized. In comparison with that, the implementation of the concept and wider adoption of these practices concerning secure software supply chains is rather limited. This section is devoted to the discussion of prior research associated with this topic in the literature in the recent past.

SLSA is a security concept aimed at providing a clear process for implementing secure practices in software supply chains [16]. SLSA introduces different tracks and levels of security to achieve within those tracks. The v1.0 of SLSA focuses on its build track and defines security levels (L0-L3), concentrating on provenance, tamper resistance, and building artifacts with hardened environments. This framework has found traction in open-source projects and CI/CD ecosystems, which allows organizations to implement strict security policies at the time of artifact building and consumption. However, integrating SLSA into tools that were not originally designed with it in mind remains operationally complex.

### *A. Tekton Chains*

Tekton Pipelines [9] is a cloud-native workflow automation system that is similar to Argo Workflows. Tekton Chains [15] was developed to become an extension of the Tekton ecosystem to enable cryptographic signing and provenance attestation for artifacts built using Tekton Pipelines. The solution is highly tethered and tailored to work within the Tekton ecosystem and does not support integration into any other external workflow automation engines. It works well for Tekton itself, but only gets up to Level 2 of the SLSA build track.

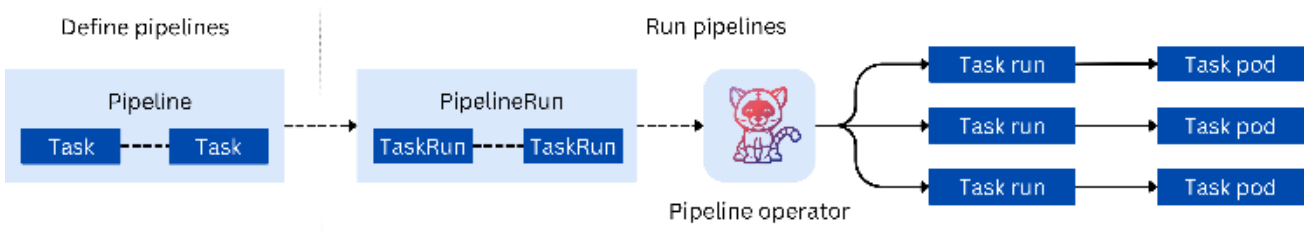

Fig. 2. Tekton Pipelines runtime architecture

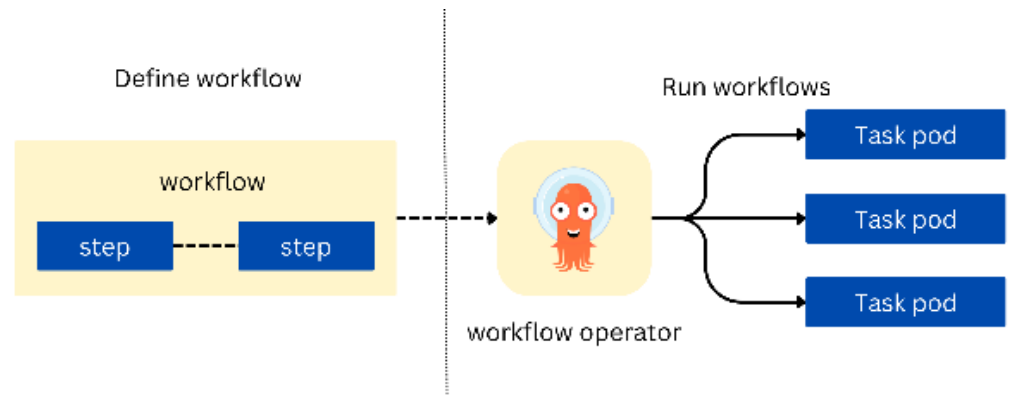

Fig. 3. Argo Workflows runtime architecture

"Fig. 2", and "Fig. 3", illustrate the distinct architectural designs of Tekton and Argo Workflows. At the time of execution, Tekton maintains a PipelineRun Custom Resource, which contains all the information about the pipeline execution and the tasks, and then TaskRuns Custom Resources to maintain information on individual tasks. So Tekton chains can work on a task level to extract information about artifacts and secure them. But when it comes to Argo Workflows at the time of execution, there will only be one Workflow Custom Resource, which will contain the information about the workflow. Because of this architecture, handling things at a task level is complex in Argo Workflows

### *B. Sigstore*

Sigstore is an open-source project that aims to ease the process of signing, verification, and transparency of cryptographic signatures for artifacts like container images, binaries, as well as source codes [19]. The main goal of the project is to make sure to enable effortless signing and validation even without user-managed cryptographic keys for the built artifacts to enhance the security of the software supply chain. Cosign is a key tool provided by Sigstore to enable effortless cryptographic signatures with or without key management. Signing without keys can be achieved by integrating the Fulcio tool provided by Sigstore. Sigstore also provides transparency by providing an immutable, tamper-resistant ledger of metadata using Rekor.

### *C. Syft and Grype*

Syft [20] is an open-source tool to generate SBOMs from container images and filesystems, and Grype [21] is its companion vulnerability scanner that analyzes those SBOMs to detect known vulnerabilities. It can scan container images, file systems, and repositories and detect security issues in operating systems and programming language dependencies.

Researchers at Montana State University assessed 2,313 Docker images by first generating SBOMs with Syft and Trivy [22], then scanning them with Grype and CVE-bin-tool [23]. Their results showed that the choice of SBOM generator influences subsequent findings: Syft's more comprehensive SBOMs led to more accurate vulnerability detection.

### *D. Kyverno*

Kyverno is a Kubernetes-native policy engine designed to enforce policies using standard YAML syntax, enabling validation, mutation, and generation of Kubernetes resources. Kyverno is useful in software supply chain security as it can verify signed container images and enforce policies based on artifact provenance, making it a critical component for policy-based governance in CI/CD workflows and runtime environments [24].

## III. Methodology

To address the gap identified within Argo Workflows, we propose a Kubernetes controller [25] that monitors and reconciles Argo Workflow custom resources. Because it operates on the same objects as the native Argo Workflows operator, the controller must perform its reconciliation logic in a non-intrusive manner to avoid conflicts. This is accomplished by consuming the open-source Argo Workflows Custom Resources, which are responsible for describing how the workflows execute within the cluster. This way of development is common when adding features to an existing system [26]. The controller acts as an extension to the Argo Workflows ecosystem.

Managing the lifecycle of a workload or a process is crucial in Kubernetes because it sits in a distributed system, and any resource can be evicted or disrupted at any given time [27]. The Argo Workflows Custom Resource is responsible for describing how the workflow should be run, and the Argo Workflows Operator is responsible for creating the pod workload to execute the desired workflow execution within the Kubernetes cluster. Because the controller proposed by this paper is built to work based on existing Argo Workflows' Custom Resources, it does not interfere with the Argo Workflows Operator.

Resources in Kubernetes are defined in a YAML configuration and stored in a key-value store called ETCD. The Kubernetes reconciliation process can be defined as an iterative process by making the desired state equivalent to the state defined in a resource configuration. During this iterative process, it is crucial to store the state of the individual resource configuration. Usually, this is done under the "status" section of a Kubernetes resource configuration. This is the case for the Argo Workflows Operator as well. If the ARGO-SLSA controller also maintains its reconciliation status under the "status" section of the workflow Custom Resource, it will lead to conflicts with the actual Argo Workflows Operator. To avoid this, the authors decided to maintain the controller statuses under the "annotations" section of the workflow resources.

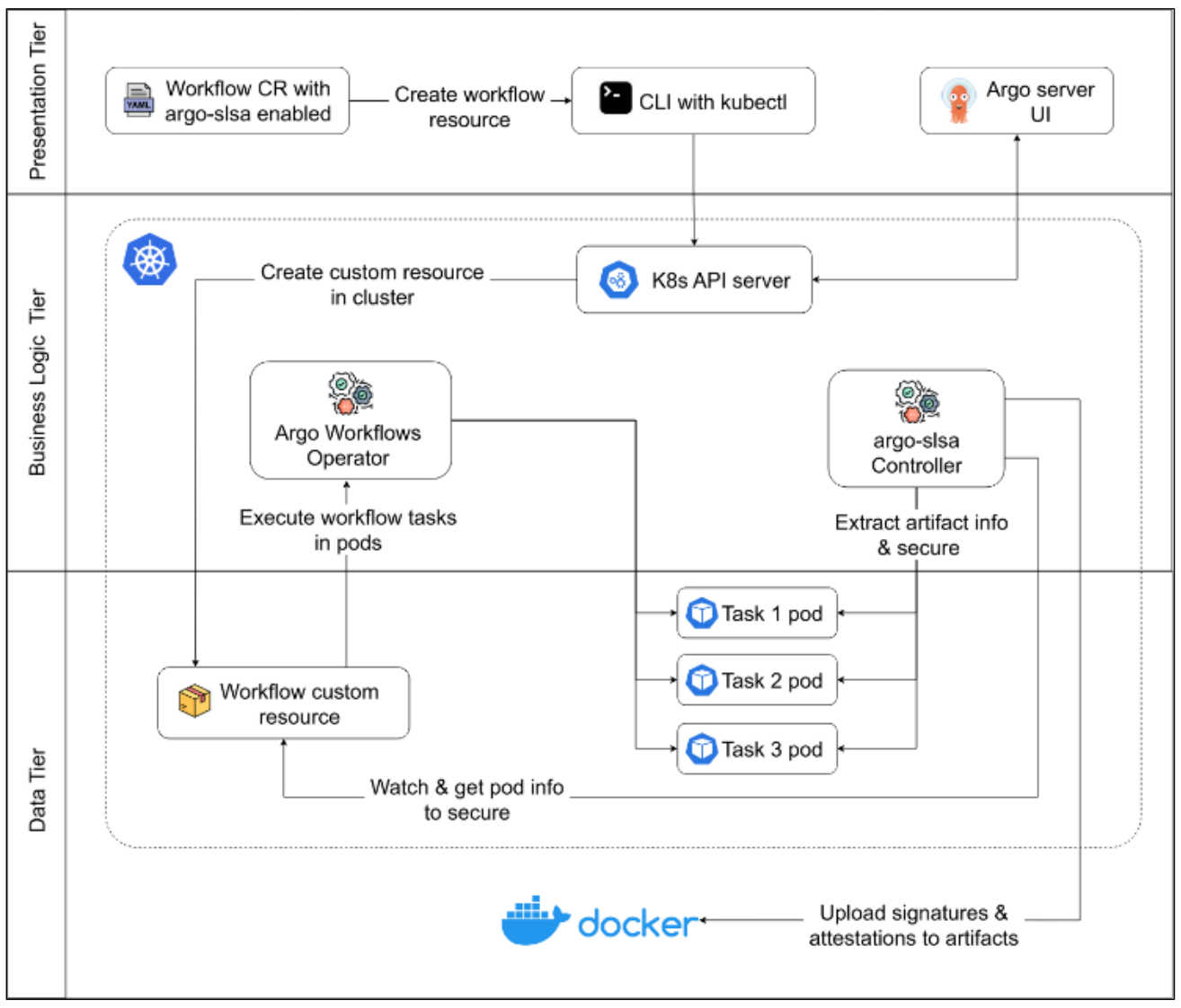

Fig. 4. ARGO-SLSA controller architecture

```
status:
  nodes:
    build-and-push-docker-zjxg9-918567359:
      boundaryID: build-and-push-docker-zjxg9
      displayName: build-and-push(1:context:/workspace/test/build-src/docker/hello-world-2,image:mothlk/hello-world-2:0.1.0)
      finishedAt: "2025-01-19T06:01:45Z"
      hostNodeName: k3d-test-cluster-server-0
      id: build-and-push-docker-zjxg9-918567359
      inputs:
        parameters:
        - name: image
          value: mothlk/hello-world-2:0.1.0
        - name: context
          value: /workspace/test/build-src/docker/hello-world-2
      name: build-and-push-docker-zjxg9[0].build-and-push(1:context:/workspace/test/build-src/docker/hello-world-2,image:mothlk/hello-world-2:0.1.0)
      outputs:
        exitCode: "0"
      phase: Succeeded
      progress: 1/1
      resourcesDuration:
        cpu: 0
        memory: 4
      startedAt: "2025-01-19T06:01:41Z"
      templateName: build-and-push
      templateScope: local/build-and-push-docker-zjxg9
      type: Pod
```

Fig. 5. Task information under the "status.nodes" section of a workflow

```
outputs:
  parameters:
    - name: OCI
      valueFrom:
        path: /tmp/oci.txt
```

```
echo -n "{{inputs.parameters.image}}@${digest}" > /tmp/oci.txt
```

Fig. 6. Exploring the artifact information

Then, this study addressed the problem of tracking individual task states within the workflow. During the runtime of a workflow, the Argo Workflows Operator spawn pods to execute the workflow steps "Fig. 4". The information on the tasks will be maintained under the "status.nodes" section of a workflow configuration by the Argo Workflows Operator "Fig. 5". With the same idea of not interfering with the Argo Workflows Operator, authors decided to maintain individual task reconciliation status in the "annotation" section of pod resources spawned for each task.

Because the implementation is made as an extension to the existing Argo Workflows, it can leverage existing features like Argo CLI and Argo server for UI. Enabling ARGO-SLSA for a workflow can be done by adding {argo.slsa.io/enable: "true"} annotation to a workflow.

To enable the ARGO-SLSA controller to identify which artifact to secure, the workflow must emit the artifact's metadata, including its exact digest. The system then uses that digest to start the process to secure the artifact. Including the digest is mandatory because it uniquely and immutably identifies the precise artifact version, eliminating ambiguity and guarding against tampering.

## IV. Results and Discussion

To evaluate the controller architecture proposed "Fig. 4", the authors have used a self-hosted Kubernetes cluster created using K3D [28] with Argo Workflows and ARGO-SLSA controller deployed with required role-based access controls (RBAC) with the least privileges. The setup and teardown of the test environment are automated with the Make [29] tool. "Fig. 6", shows the available commands related to the test environment context. The evaluations are done manually, triggering workflows to check if the controller behaves as intended. In addition to the manual tests, automated unit tests using the Golang Test Suite have been implemented to evaluate the controller functionalities. "Fig. 7", shows the test coverage of the controller.

```
Environment
  setup-env          Setup the environment with k3d to run the controller.
  teardown-env       Teardown the k3d environment.
  stop-env           Stop the k3d environment.
  start-env          Start the k3d environment.
  import-image-to-k3d  Import the controller image to the k3d cluster.
  build-and-deploy-controller  Build and deploy the controller to the k3d cluster.
  remove-controller  Remove the controller from the k3d cluster
```

Fig. 7. Automated environment setup and teardown using Make

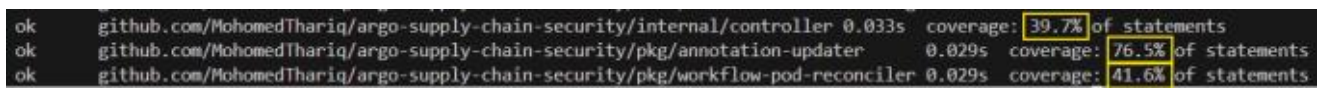

Fig. 8. ARGO-SLSA controller unit tests coverage

Following are the tests conducted against the system to evaluate and ensure it meets the research objectives.

### A. Providing insights: Using available tooling in the Argo Workflows ecosystem to provide feedback ARGO-SLSA controller Operations

Since the ARGO-SLSA controller extends the existing Argo Workflows custom resource definition (CRD), it should be able to handle any workflow initializations by any supported mechanism (TABLE I). Beyond its trigger mechanisms, the controller must also integrate with available tools to report progress and improve overall usability.

TABLE I. SUPPORT FOR TRIGGER METHODS AND FEEDBACK

| Trigger Method | Trigger | Feedback |
|---|---|---|
| Using “kubectl” | ☑ | ☑ |
| Using ArgoCLI | ☑ | ☒ |
| Using Argo Server UI | ☑ | ☑ |

### B. Feature toggle: Make sure the controller can identify feature-enabled workflows and ignore not-enabled workflows

The ARGO-SLSA controller is designed as a plugin to the existing Argo Workflows ecosystem. In case a user does not need the feature to be enabled for a workflow, the user should be able to exclude it from the ARGO-SLSA controller. Because of this, authors have added a feature toggle to only enable when needed. The feature is enabled by adding {argo.slsa.io/enable: "true"} annotation to the workflow resource. “Fig. 9”, shows how the controller only executes its logic against workflows with {argo.slsa.io/enable: "true"} annotation. This corresponds to proving **RO1**.

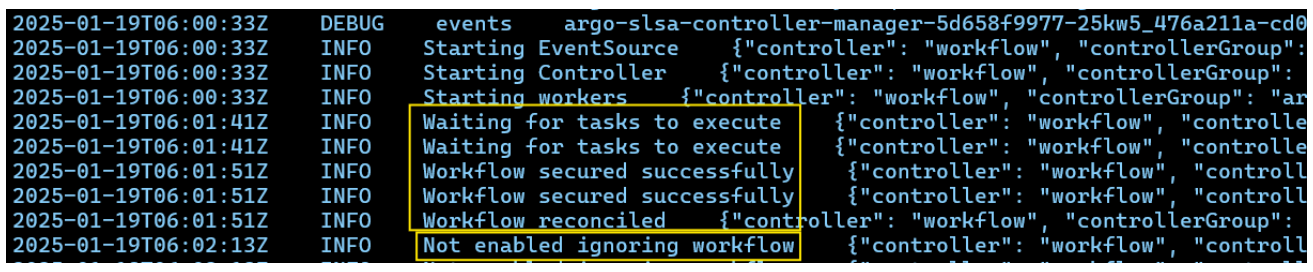

Fig. 9. Feature toggle

### C. High availability: Make sure the controller is implemented to support high availability

The underlying controller uses Operator-SDK, which uses Kubebuilder [30] for Golang-based controllers. This provides most of the scaffolding to implement a Kubernetes controller comparatively fast. Given that Kubernetes is a distributed system, the workloads can be evicted or disrupted at any time. This reason motivated the implementation of things to work with high availability. Here, Kubebuilder also provides leader election functionality to controllers built with it. Once the controller is deployed into the cluster with multiple replicas controller will have the intelligence to elect a leader among the replicas to do the reconciliation on workflows. In case the lead controller gets evicted, the remaining replicas will assign a leader among them to continue the reconciliation process.

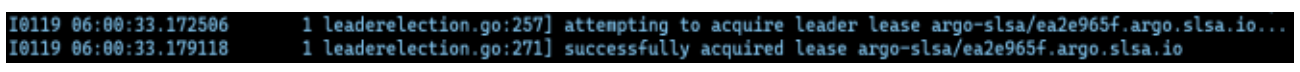

Fig. 10. Controller pod replicas leader election

### D. Maintain status: Make sure the controller maintains its status of workflows and its tasks, so that in case of a controller pod eviction, it can pick up where it left off

Unlike Tekton Pipelines, Argo Workflows do not have their own Custom Resources for task-level objects. Since ARGO-SLSA is a plugin controller to Argo Workflows; it will not own the workflow Custom Resource and will have to work with the Argo Workflows Operator. Because of this, if the controller directly maintains its status in the resource’s status section, it can lead to conflicts with the Argo Workflows Operator. This is the reason the status is updated in the “annotations” section of the resource. When it comes to individual task statuses, the pod annotations section will be used. This shows that **RO2** can be accessed.

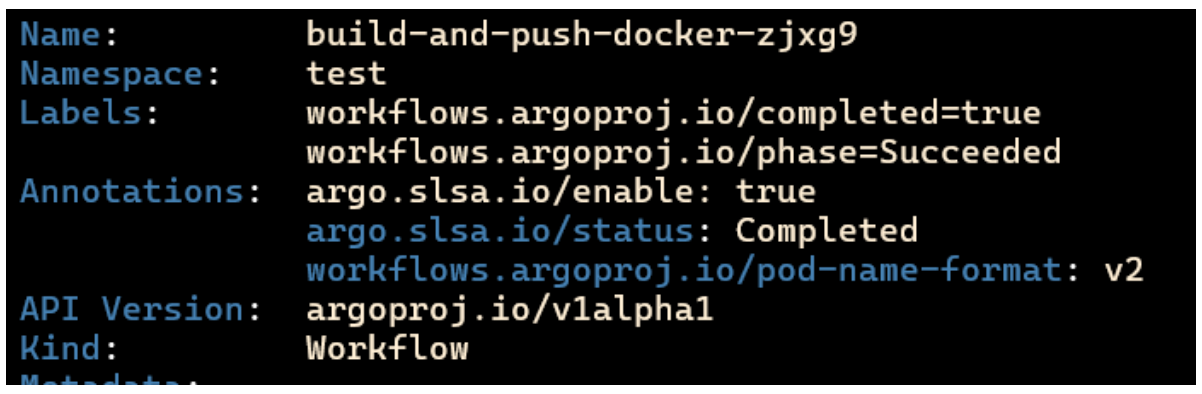

Fig. 11. How status is maintained in a Workflow resource

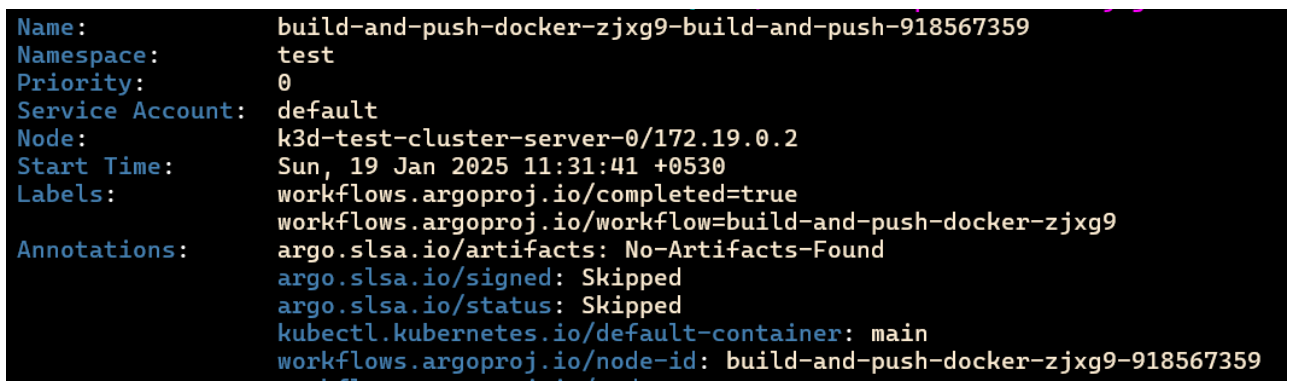

Fig. 12. How status is maintained in a pod – Workflow task

### E. SLSA in Argo Workflows: Make sure to provide supply chain security to artifacts built using Argo Workflows

The implementation for SLSA enforcement can be verified by using cosign to validate final artifact signatures and attestation. This answers **RO3**.

```
# Command to verify OCI artifact signature
cosign verify <OCI-Image> \
  --certificate-identity "https://kubernetes.io/
      namespaces/<argo-slsa-namespace>/
      serviceaccounts/<serviceaccount-of-controller>
      " \
  --certificate-oidc-issuer <cluster-Open-ID-Issuer>

# Command to verify OCI artifact attestation
cosign verify-attestation --type slsaprovenance <OCI
    -Image> \
  --certificate-identity "https://kubernetes.io/
      namespaces/<argo-slsa-namespace>/
      serviceaccounts/<serviceaccount-of-controller>
      " \
  --certificate-oidc-issuer <cluster-Open-ID-Issuer>
```

Fig. 13. Verify OCI artifact signature and attestation

### F. Address dependency attacks: Make sure external dependencies are pinned to the artifact as SBOM.

The controller has Syft and Grype integrated into the system to generate and analyze SBOMs for each image created for specific architectures. This approach enables organizations to enhance software supply chain security across various artifact types. Tools like Grype allow consumers to scan for known threats that might be available in the artifact consumed.

```
# Verify SBOM for a specific image architecture
cosign verify-attestation --type cyclonedx <OCI-Image> \
  --certificate-identity https://kubernetes.io/namespaces/<argo-slsa-namespace>/serviceaccounts/<serviceaccount-of-controller> \
  --certificate-oidc-issuer <cluster-Open-ID-Issuer> | jq -r '.payload' | base64 -d | jq -r '.predicate' > sbom.json

# Check vulnerabilities of the image
grype sbom:./sbom.json
```

Fig. 14. Vulnerability scanning based on a signed SBOM attested by the system

### G. *Enforce deployment integrity: Use of Kyverno to block unverified artifacts during deployments.*

Once software artifacts are cryptographically signed and include provenance attestations, Kyverno [24] is used as a policy enforcement engine that validates artifacts at runtime. Using Kyverno's "verifyImages" policy, administrators mandate that only container images signed, accompanied by valid SLSA-compliant attestations, are admitted into the cluster.

Thereafter, an evaluation between Tekton Chains and the ARGO-SLSA controller was carried out (TABLE II). Overall, the ARGO-SLSA controller delivers software-supply-chain security to any workflow that produces supported artifact types. Its ability to generate SBOMs with multi-architecture support gives both producers and consumers clear visibility into upstream dependencies and enables early threat scanning, thereby identifying and mitigating vulnerabilities much earlier in the development lifecycle. The ARGO-SLSA controller's explicit state management enhances system reliability. In Tekton Chains, any controller failure retriggers a full reconciliation restart, which can produce duplicate signatures and attestations for the same artifact and introduce ambiguity into the verification process. With explicit state handling, the ARGO-SLSA controller continues the process where it left over, providing better disaster recovery.

TABLE II. EVALUATION BETWEEN TEKTON CHAINS VS ARGO-SLSA

| Criteria | Tekton Chains | ARGO-SLSA |
|---|---|---|
| Secure OCI artifacts | ☑ | ☑ |
| Secure other artifact types | ☒ | ☒ |
| Sign Identified artifacts | ☑ | ☑ |
| Attach SLSA compliant provenance to the artifacts | ☑ | ☑ |
| Attach SBOM to Identify external dependencies | ☒ | ☑ |
| Explicit state handling to continue reconciliation from the stopped point in case of a controller pod eviction | ☒ | ☑ |

## V. CONCLUSION AND FUTURE WORK

In this paper, the authors investigate how to improve software supply chain security for artifacts built through Argo Workflows. The evaluation confirms that the ARGO-SLSA controller inspects workflow resources and their constituent tasks, signs produced artifacts, generates standardized SBOMs, and attaches cryptographic attestations at precise digest values to ensure Level 2 of the SLSA compliance build track. Thereby, bridge the vacuum in software supply chain security aspects such as provenance tracking and artifact tamper resistance, and enable early detection of potential threats.

The current implementation focuses on the Open Container Initiative (OCI) artifact format, for container images & helm charts [31]. Thereby in future this could be extended to broader artifact types, and provide support to different storage backends to publish signature, provenance and SBOM information for those artifact types. Additionally, the controller is specifically designed for the Argo Workflows ecosystem. Which limits its applicability to other workflow automation platforms like Tekton Pipelines [32], Jenkins X [33], or OpenShift Pipelines [34] without substantial reengineering. Therefore, opportunities remain for designing a universal architecture that attains SLSA Level 3 compliance.

## VI. SOURCE CODE

The source code for the proposed solution is available at https://github.com/MohomedThariq/argo-supply-chain-security